\documentclass[11pt,a4paper]{article}
\usepackage{jheppub}

\usepackage{amsmath}
\usepackage{amsfonts}
\usepackage{slashed}
\usepackage{graphicx}
\usepackage{mathrsfs}
\usepackage{axodraw}
\usepackage[numbers,sort&compress]{natbib}

\makeatletter
\g@addto@macro\bfseries{\boldmath}
\makeatother

\newcommand{\eps}{\epsilon}
\newcommand{\ord}{\begin{cal}O\end{cal}}

\def\beq{\begin{equation}}
\def\eeq{\end{equation}}
\def\bsp#1\esp{\begin{split}#1\end{split}}

\makeatletter
\newcommand{\IEIF}{%
  \def\@IEIFsep{(}%
  I_F\@IEIFi
}
\newcommand\@IEIFi{\@ifnextchar\stopIEIF{\@IEIFend}{\@IEIFii}}
\newcommand\@IEIFii[4]{%
  \big\@IEIFsep
  \begin{smallmatrix}
    #1 & #2 \\
    #3 & #4
  \end{smallmatrix}
  \def\@IEIFsep{|}
  \@IEIFi
}
\newcommand\@IEIFend[2]{%
  ; #2 \bigr)
}
\makeatother

\newcommand{\cE}{\begin{cal}E\end{cal}}

\newcommand{\cQ}{\begin{cal}Q\end{cal}}

\newcommand{\cEf}[4]{{\mathcal{E}_4}\!\left(\begin{smallmatrix}#1\\#2\end{smallmatrix};#3,#4\right)}

\renewcommand{\ln}{\log}

\newcommand{\bx}{{\bf x}}

\title{Analytic results for two-loop planar master integrals for Bhabha scattering} 

\author[a]{Claude Duhr}
\author[b,c]{Vladimir A.\ Smirnov} 
\author[d]{Lorenzo Tancredi}
\affiliation[a]{Bethe Center for Theoretical Physics, Universit\"at Bonn, D-53115, Germany}
\affiliation[b]{Skobeltsyn Institute of Nuclear Physics of Moscow State University,
Moscow 119992, Russian Federation}
\affiliation[c]{Moscow Center for Fundamental and Applied Mathematics, Moscow 119992, Russian Federation}
\affiliation[d]{Rudolf Peierls Centre for Theoretical Physics, University of Oxford,
Clarendon Laboratory, Parks Road, Oxford OX1 3PU}

\emailAdd{cduhr@uni-bonn.de}
\emailAdd {smirnov@theory.sinp.msu.ru}
\emailAdd{lorenzo.tancredi@cern.ch}

\abstract{We analytically evaluate the master integrals for the second type of planar contributions 
to the massive two-loop Bhabha scattering in QED using differential equations with canonical bases.
We obtain results in terms of multiple polylogarithms for all the master integrals but one, for which we derive a compact result in terms of elliptic multiple polylogarithms. As a byproduct, we also provide a compact analytic result in terms of elliptic multiple polylogarithms 
for an integral belonging to the first family of planar Bhabha integrals, whose computation in terms of polylogarithms
was addressed previously in the literature. }


\begin{document}

\rightline{
BONN-TH-2021-06\,, 
OUTP-21-19P
}

\maketitle

\catcode`\@=11
\font\manfnt=manfnt
\def\Watchout{\@ifnextchar [{\W@tchout}{\W@tchout[1]}}
\def\W@tchout[#1]{{\manfnt\@tempcnta#1\relax%
  \@whilenum\@tempcnta>\z@\do{%
    \char"7F\hskip 0.3em\advance\@tempcnta\m@ne}}}
\let\foo\W@tchout
\def\dubious{\@ifnextchar[{\@dubious}{\@dubious[1]}}
\let\enddubious\endlist
\def\@dubious[#1]{%
  \setbox\@tempboxa\hbox{\@W@tchout#1}
  \@tempdima\wd\@tempboxa
  \list{}{\leftmargin\@tempdima}\item[\hbox to 0pt{\hss\@W@tchout#1}]}
\def\@W@tchout#1{\W@tchout[#1]}
\catcode`\@=12


\section{Introduction}
\label{sec:introduction}

The Bhabha scattering process, i.e., the elastic scattering of an electron-positron pair, is one of the standard candles at lepton colliders, and it will  play a crucial role at future circular or linear colliders. A precise theoretical knowledge of this process, including up to next-to-next-leading order effects in the QED coupling constant, is therefore highly desirable. So far the NNLO cross section is only known in the massless limit, supplemented by the leading logarithmic finite mass effects (see, e.g., ref.~\cite{Banerjee:2021mty} for recent results). Complete NNLO results including the full dependence on the electron mass, however, are not yet available.  

One of the main obstacles to obtain the complete NNLO results is the complexity of the two-loop integrals involved. The relevant two-loop integrals were evaluated in the small-mass limit in 
refs.~\cite{Actis:2006dj,Actis:2007gi,Actis:2007pn,Actis:2007fs,Actis:2008br}. Up to now, there are only partial results for two-loop integrals with massive fermions. 
Analytic results for diagrams with one-loop insertions and a closed massive fermion loop were obtained in refs.~\cite{Bonciani:2003cj,Bonciani:2004gi}. 
First analytic results for massive two-loop 
double-box Bhabha diagrams were obtained in refs.~\cite{Smirnov:2001cm, Heinrich:2004iq}. More attempts to evaluate two-loop Bhabha integrals can be found in 
refs.~\cite{Czakon:2004wm,Czakon:2005jd,Czakon:2005gi,Czakon:2006pa,Czakon:2006hb}. 

\begin{figure}[h]
\begin{center}
\includegraphics[scale=1.]{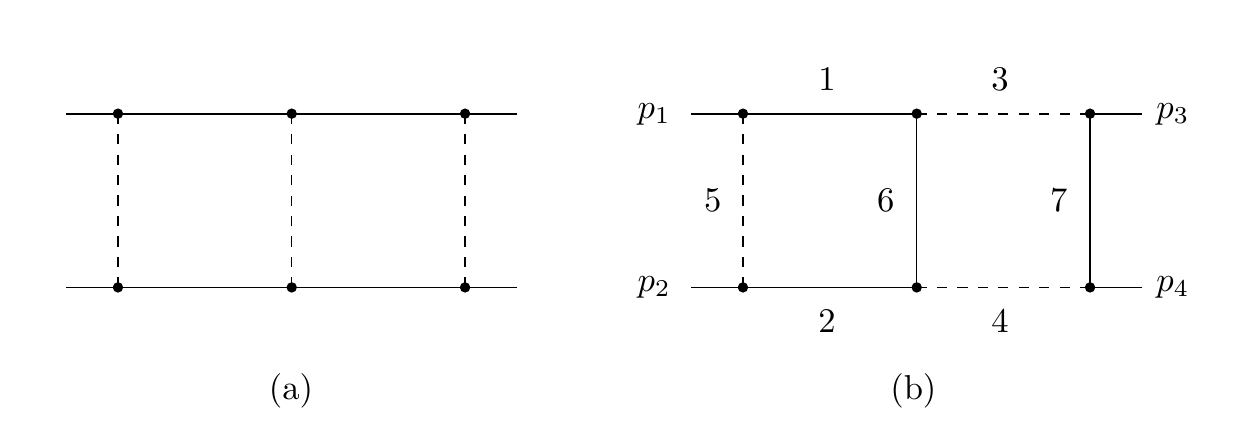}
\caption{Planar graphs of the first and the second type for two-loop Bhabha scattering. 
Solid (dashed) lines indicate massive (massless) propagators. All the external momenta are incoming.}
\label{plaBha}       
\end{center}
\end{figure}

The master integrals relevant for the calculation of the two-loop corrections to Bhabha scattering can be classified into two planar families and one non-planar one.
The systematic evaluation of the integrals in the first family was started in ref.~\cite{Henn:2013woa},
 where the master integrals 
for the family associated with graph (a) of fig.~\ref{plaBha} were evaluated in terms of multiple 
polylogarithms~\cite{Lappo:1927,Goncharov:1998kja,GoncharovMixedTate} (MPLs)
in the framework of the method of differential equations \cite{Kotikov:1990kg,Remiddi:1997ny,Gehrmann:1999as} 
with the help of the strategy  based on canonical bases~\cite{Henn:2013pwa}.
In ref.~\cite{Henn:2013woa}, a solution in terms of MPLs had been provided for all integrals except one,  whose analytic evaluation
was hindered by the presence of the a non-rationalisable square-root in the symbol alphabet. More recently, it was shown that
also this last integral can be expressed in terms of multiple polylogarithms by an integration technique based
on an ansatz of MPLs with suitable arguments~\cite{Heller:2019gkq}.

\begin{figure}[h]
\begin{center}
\includegraphics[scale=1.]{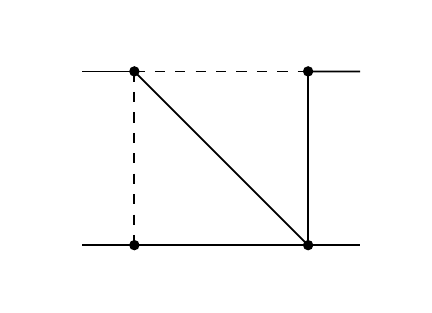}
\caption{The graph associated with the master integral which we evaluate in terms of elliptic polylogarithms.}
\label{el14}       
\end{center}
\end{figure}

The goal of the present paper is to
analytically evaluate the master integrals for the second planar family, which is associated with graph (b) of fig.~\ref{plaBha}. 
In a way that is reminiscent of the first planar family of integrals, we will show that also in this case 
we can obtain results in terms of MPLs for all  master integrals but one (see fig.~\ref{el14}), 
due to the presence of the same non-rationalisable
square root found in the evaluation of the master integrals for graph (a).
While it can be shown by direct integration techniques that also for this master integral a representation in terms of MPLs exist, 
the representation we obtained is extremely cumbersome and of no practical use. 
Nevertheless, it turns out that a compact result for this integral can be derived in terms of the elliptic MPLs introduced in refs.~\cite{Broedel:2017kkb,Broedel:2017siw,Broedel:2018iwv}.
As a byproduct of our analysis, we will also
provide a very compact analytic result for the remaining master integral in the first family in fig.~\ref{plaBha}(a) in terms
of the same class of functions.\footnote{This result had first been presented by one of the authors at the \emph{Loops and Legs Conference 2018} in St. Goar, but it has never been published before.} 

The rest of the paper is organised as follows. In section~\ref{sec:can_diff_eqs} we explain the notation, 
present the system of canonical differential equations for the problem at hand, and introduce the alphabet needed to describe the planar family in fig.~\ref{plaBha}(b). The alphabet contains multiple square roots, and it is a priori not
obvious that the differential equations can be integrated in terms of multiple polylogarithms.
In section~\ref{sec:mpls} we  review general sufficient criteria and algorithms to solve a system of canonical differential equations in terms of MPLs or their elliptic generalisations.
We then
continue in section~\ref{sec:can_diff_eqs_part2} to apply these ideas to our computation.
We explain   
how our differential equations can straightforwardly be solved in terms of multiple polylogarithms 
for all elements of the basis of the master integrals but one,
which is connected with the graph shown in fig.~\ref{el14}.
In section~\ref{sec:f14} we show how a convenient and compact representation
for this integral can be found in terms of elliptic generalisations of MPLs. 
We also provide a very compact, alternative, analytic solution for one of the 
master integrals in the first planar family considered in refs.~\cite{Henn:2013pwa,Heller:2019gkq}.
Finally we draw our conclusions in section~\ref{sec:conclusions}.


\section{Canonical differential equations}
\label{sec:can_diff_eqs}

The Feynman integrals for the family of fig.~\ref{plaBha} (b) can be organised in an integral family with nine propagators, 
where the first seven propagators correspond to the edges of the graph and the last two are so-called irreducible numerators,
\beq\bsp\label{eq:Bhabha_family}
&F_{a_1,a_2,\ldots,a_9}(s,t,m^2;D) = \left(\frac{e^{\gamma_E\eps}}{i \pi^{D/2}}\right)^2
\int\frac{d^Dk_1\,d^Dk_2}{[-k_1^2+m^2]^{a_1}[-(k_1+p_1+p_2)^2+m^2]^{a_2}[-k_2^2]^{a_3}}\\
&\quad\times\frac{[-(k_2+p_1)^2]^{-a_8}\,[-(k_1-p_3)^2]^{-a_9}}{[-(k_2+p_1+p_2)^2]^{a_4}
[-(k_1+p_1)^2]^{a_5}[-(k_1-k_2)^2+m^2]^{a_6}[-(k_2-p_3)^2+m^2]^{a_7}}\,.
\esp\eeq
The indices $a_i$, $i=1\ldots7$ can be positive or negative integers, but we restrict our computation to $a_8,a_9 \leq 0$. 
We work in dimensional regularisation in $D=4-2\eps$ dimensions in order to regulate
both infrared and ultraviolet divergences. 
The electron mass is denoted by $m$, and the external momenta $p_i$ are on shell, $p_i^2=m^2$. We introduce the usual Mandelstam variables
\beq \label{eq:mandelstam}
s=(p_1+p_2)^2\,, \qquad t=(p_1+p_3)^2\,, \qquad u=(p_2+p_3)^2\,,
\eeq
with $s+t+u=4m^2$.

Using the public codes {\tt FIRE}~\cite{Smirnov:2019qkx} and {\tt KIRA}~\cite{Maierhoefer:2017hyi,Klappert:2020nbg}, we solve the 
integration-by-parts (IBP)  identities~\cite{Chetyrkin:1981qh,Tkachov:1981wb} and reveal 43 independent master integrals, which we collect into the vector
$g=(g_1,\ldots,g_{43})^T$. This vector satisfies a system of linear differential equations of the form
\beq
\label{eq:DEs}
\partial_v g = A_v g\,,
\eeq
where $v=s,t,m^2$, $\partial_v=\frac{\partial}{\partial v}$ and the matrices $A_s,A_t,A_{m^2}$ are rational functions of
$s,t,m^2$ and $\epsilon$. In the following, it will be useful to collect all the partial derivatives into a total differential, 
and to work with the equation
\beq
\label{eq:DEs2}
d g = dA g\,,\qquad dA = ds\,A_s+dt\,A_t+dm^2\,A_{m^2}\,.
\eeq

In order to evaluate the master integrals, it is convenient to search for a so-called  canonical basis~\cite{Henn:2013pwa}, i.e.,
a basis of master integrals for which the matrix $dA$ on the right-hand side of eq.~(\ref{eq:DEs2}) is proportional to $\epsilon$ and its entries can
all be expressed as linear combinations of total differentials of logarithms.
While it has been suggested (at least conjecturally) that the study of the residues of the integrands can provide the full information to determine the elements of the
canonical basis~\cite{ArkaniHamed:2010gh}, 
this analysis can become computational very expensive when square-roots are involved.\footnote{See ref.~\cite{Henn:2020lye} for an automated implementation of this approach.} Therefore, we
follow here a mixed approach to find a canonical basis. In order to select suitable candidates, we start by analysing only the leading singularities associated to the maximal cuts of the various integrals, and we supplement this analysis by
the  method of ref.~\cite{Gehrmann:2014bfa}. 
By choosing integrals with unit leading singularities at the level of the maximal cuts, 
one can often bring the initial differential equations
into a so-called precanonical form, where the corresponding matrices depend linearly  on $\eps$.
Once this is achieved, the prescriptions of ref.~\cite{Gehrmann:2014bfa} can be successfully applied to arrive at a fully canonical form. 
In our case, the precanonical basis is composed of the $43$ Feynman integrals present on the right-hand side of the 
above expression for the $\eps$-basis. 
In this way we obtain the new system of differential equations
\begin{equation} \label{eq:DEs-ep2}
df = \eps\, d\bar{A}\, f\,,
\end{equation}
with $\bar{A}$ independent of $\epsilon$. 
Our canonical basis $f_i$ is given in appendix~\ref{app:canonical_basis}.
The price to pay for casting the equations in  canonical form is that
the new matrix $d\bar{A}$ is not a matrix of rational one-forms, but it involves four square roots:
\begin{align}
\label{eq:sqrts}
r_s  &\,=  \sqrt{-s} \sqrt{4 m^2-s},\qquad r_t = \sqrt{-t}\sqrt{4 m^2-t}, \\ 
\nonumber r_u &= \sqrt{-s-t} \sqrt{4 m^2-s-t}, 
\quad
 r_{st} \, =  \sqrt{-s} \sqrt{4 m^6-s (m^2-t)^2} \,.
\end{align}
More precisely, the matrix $\bar{A}$ takes the general form
\begin{align}
\bar{A} = \sum_{i=1}^{16}\bar{A}_i\,\log R_i(s,t,m^2)\,, \label{eq:Abar_def}
\end{align}
where the $R_i$ are algebraic functions, referred to as \emph{letters} (and their collection is called the \emph{alphabet}):
\begin{align}
\nonumber R_1(s,t,m^2)&\,=\frac{s}{m^2}\,,\qquad R_2(s,t,m^2) =\frac{s-4 m^2}{m^2}\,, \qquad R_3(s,t,m^2)=\frac{s-r_s}{s+r_s}\,,\\
\nonumber R_4(s,t,m^2)&\,=\frac{t}{m^2}\,,\qquad R_5(s,t,m^2)=\frac{t-4 m^2}{m^2}\,,\qquad R_6(s,t,m^2)=\frac{t-r_t}{t+r_t}\,,\\
\nonumber R_7(s,t,m^2)&\,=\frac{s+t}{m^2}\,,\qquad R_8(s,t,m^2)=\frac{s+t-4 m^2}{m^2}\,,\qquad R_9(s,t,m^2)=\frac{s+t-r_u}{s+t+r_u}\,,\\
\label{eq:R_letters}
R_{10}(s,t,m^2)&\,=\frac{st+r_s r_t}{st-r_s r_t}\,,\qquad R_{11}(s,t,m^2)=\frac{(s+t) r_s-s r_u}{(s+t) r_s+s r_u}\,,\\
\nonumber R_{12}(s,t,m^2)&\,=\frac{(s+t) r_t-t r_u}{(s+t) r_t+t r_u}\,,\qquad R_{13}(s,t,m^2)=\frac{r_{st}-m^2 s+s t}{r_{st}+m^2 s-s t}\,,\\
\nonumber R_{14}(s,t,m^2)&\,=\frac{r_t r_{st}-3 m^2 s t+s t^2}{r_t r_{st}+3 m^2 s t-s t^2}\,,\qquad R_{15}(s,t,m^2)=\frac{r_s r_{st}+4 m^4 s-m^2 s^2+s^2 t}{r_s r_{st}-4 m^4 s+m^2 s^2-s^2 t}\,,\\
\nonumber R_{16}(s,t,m^2)&\,=\frac{m^2 r_s r_{st}-s r_s r_{st}-4 m^6 s-3 m^4 s^2+m^2 s^3+3 m^2 s^2 t-s^3 t}{m^2 r_s r_{st}-s r_s r_{st}+4 m^6 s+3 m^4 s^2-m^2 s^3-3 m^2 s^2 t+s^3 t}\,.
\end{align}
As we will see, the appearance of these square roots makes the solution of the differential equation~\eqref{eq:DEs-ep2} in terms of MPLs highly non-trivial.

It is easy to write down a formal solution to eq.~\eqref{eq:DEs-ep2} as 
\beq\label{eq:g_path_sol}
 f(s,t,m^2;\eps) = \mathbb{P}\textrm{exp}\left[\eps\int_{\gamma}d\tilde{A}\right]\,f_0(\eps)\,,
\eeq
where $\mathbb{P}\textrm{exp}$ denotes the path-ordered exponential and $f_0(\eps)$ encodes the initial condition and is related to the value of $f$ at a specific point $(s_0,t_0,m_0^2)$. The path $\gamma$ connects the initial point $(s_0,t_0,m_0^2)$ to the generic point $(s,t,m^2)$. 

When expanding eq.~\eqref{eq:g_path_sol} in $\eps$, then at each order we can write $f$ in terms of Chen iterated integrals~\cite{ChenSymbol}, defined in the following way: consider a path $\gamma$ and a collection of one-forms $\omega_i$. If $t$ denotes a local coordinate on $\gamma$, we can pull each $\omega_i$ back to $\gamma$ and write $\gamma^*\omega_i = dt\,F_i(t)$. We can then define the iterated integral of a sequence $\omega_{i_1}\ldots\omega_{i_n}$ (usually referred to as a \emph{word}) by
\beq
\int_{\gamma}\omega_{i_1}\ldots\omega_{i_n} = \int_{0\le t_1\le\ldots\le t_n\le1}dt_n\,F_{i_n}(t_n)\int_0^{t_n}\ldots\int_0^{t_2}dt_1\,F_{i_1}(t_1)\,.
\eeq
The number $n$ of integrations is called the {length} of the iterated integral.
In general, this integral will not be homotopy-invariant, i.e., it will depend on the details of the path $\gamma$. There is a necessary and sufficient condition, called the {integrability condition}, for a combination of iterated integrals to be homotopy-invariant~\cite{ChenSymbol}. The details of this criterion are not important in the following. Here it suffices to say that it is always satisfied for the solutions in eq.~\eqref{eq:g_path_sol}. The corresponding Chen iterated integrals are then multi-valued functions of the end point $(s,t,m^2)$ of the path $\gamma$, where the multi-valuedness only comes from choosing two non-homotopic paths from $(s_0,t_0,m_0^2)$ to $(s,t,m^2)$.

The master integrals $f(s,t,m^2;\eps)$ can be expressed at every order in terms of Chen iterated integrals over words from the alphabet in eq.~\eqref{eq:R_letters}. The coefficient of $\eps^n$ in the path-ordered exponential in eq.~\eqref{eq:g_path_sol} only involves iterated integrals of length $n$. Chen iterated integrals are a very general class of functions, and it can be useful to express them in terms of a class of functions that are well-studied in the literature.
In the next section we discuss sufficient criteria for when Chen iterated integrals over $d\log$-forms with algebraic arguments,
as the ones above, can be expressed in terms of other classes of functions.



\section{From $d\log$-forms to multiple polylogarithms} 
\label{sec:mpls}

Since the solution of differential equations in canonical form
as Laurent series in dimensional regularisation
 leads naturally  to linear combinations of Chen iterated integrals over words of $d\log$-forms, it is natural to ask when it possible to evaluate such  iterated integrals in terms of other classes of functions, and if it is possible to perform this rewriting algorithmically. The advantage of this rewriting lies in the fact that these classes of functions may be well studied in the literature, and there may be established methods or computer codes for their manipulation and/or their numerical evaluation.

Let us consider Chen iterated integrals of the form $\int_{\gamma}w$, where $\gamma$ is a path from an initial point ${\bf x}_0 = (x_{0,1},\ldots,x_{0,p})$ to the point ${\bf x}_1=(x_{1,1},\ldots,x_{1,p})$, and $w$ is an integrable combination of words of length $n$ of the form
\beq\label{eq:generic_word}
 w= \sum_{{i_1\ldots i_n}} c_{i_1\ldots i_n}\,d\log R_{i_1}({\bf x})\ldots d\log R_{i_n}({\bf x})\,,\qquad c_{i_1\ldots i_n}\in\mathbb{Q}\,.
\eeq
We consider the initial point ${\bf x}_0$ fixed, and we see the integral as a multi-valued function of the end-point ${\bf x}_1$.
The arguments of the logarithms are assumed to be algebraic functions of the variables ${\bf x} = (x_1,\ldots,x_p)$.

For some time, there was a folklore belief in the physics community that all such iterated integrals could 
be evaluated in terms of a rather simple class of iterated integrals, namely the so-called multiple polylogarithms (MPLs), 
defined by~\cite{Lappo:1927,Goncharov:1998kja,Remiddi:1999ew,GoncharovMixedTate}
 \beq\label{eq:Mult_PolyLog_def}
 G_{a_1,\ldots,a_n}(z)= G(a_1,\ldots,a_n;z)=\,\int_0^z\,\frac{d t}{t-a_1}\,G(a_2,\ldots,a_n;t)\,,
\eeq
 where the recursion starts with $G(;z)\equiv 1$. In the special case where all the $a_i$'s are zero, we define
\beq\label{eq:GLog}
G_{0,\ldots,0}(z) = G(\underbrace{0,\ldots,0}_{n};z) = \frac{1}{n!}\,\ln^n z\,.
\eeq
The shared belief in physics was that, whenever the $d\log$'s in eq.~\eqref{eq:generic_word} have algebraic arguments
 $R_i$, then this integral can be written in terms of MPLs whose arguments $a_i$ and $z$ are algebraic functions of 
 ${\bf x}_0$ and ${\bf x}_1$. This belief was shown to be false in ref.~\cite{Brown:2020rda}, where an example of an iterated
  integral of length two was constructed that cannot be evaluated in terms of MPLs with algebraic arguments. 
 The result of ref.~\cite{Brown:2020rda} shows that the question of whether  an iterated integral of $d\log$-forms with 
 algebraic arguments can be expressed in terms of MPLs depends in general on the details of the integral, 
 including the details of the integration contour.

In the remainder of this section we discuss some special cases of algebraic letters $R_i$ where the 
Chen iterated integrals can be evaluated in terms of other classes of special functions. The starting point is the observation that in the context of Feynman integrals, the $R_i$ are  not generic algebraic functions, but often  involve at most square roots of polynomials, i.e., they are of the form:
\beq\label{eq:general_Ri}
R_i({\bf x}) = r_{i,0}({\bf x}) + \sum_{a=1}^{N_{\textrm{sqrt}}}r_{i,a}({\bf x})\,\sqrt{q_a({\bf x})}\,,
\eeq
where the $r_{i,j}$ are rational functions, and the $q_a$ are polynomials. Without loss of generality, we assume 
the $r_{i,j}$ to be polynomials.  At this point we have to make a comment: The integrand of the iterated integration, and thus the functional form of the letters $R_i$, is sensitive to a change of variables. In particular, one may ask if we can find a change of variables ${\bf x} = \psi({\bf y})$, with $\psi$ a rational function, such that $q_a(\psi({\bf y})) = u_a({\bf y})^2$ is a perfect square, for some values of $a$. This operation is known as \emph{rationalisation of square roots}. Over the last few years, several algorithmic criteria have been developed to find such a parametrisation for specific Feynman integral computations, 
or to prove that rationalisation is instead not possible~\cite{Festi1,Festi:2018qip,Besier:2018jen,Besier:2019kco,Besier:2020klg,Besier:2020hjf,Festi:2021tyq}. Note that rationalisability can easily be decided for a single square root of a one-variable polynomial ($p=N_{\textrm{sqrt}}=1$) based on the degree $\deg q_1$ of the polynomial: the square root $\sqrt{q_1(x_1)}$ can be rationalised (and the corresponding function $\psi$ can be constructed explicitly) if and only if $\deg q_1\le 2$. The question of the rationalisability for $p>1$ is much more involved, even in the presence of a single square root, see refs.~\cite{Besier:2018jen,Besier:2019kco,Besier:2020klg,Besier:2020hjf,Festi:2021tyq}.\\

In the following we discuss two special cases of eq.~\eqref{eq:general_Ri}, 
in which we can express the Chen iterated integrals in terms of other classes of iterated integrals: 
\begin{itemize}
\item If $N_{\textrm{sqrt}}=0$, the Chen iterated integrals can be expressed in terms of MPLs evaluated at algebraic arguments.
\item If $N_{\textrm{sqrt}}=1$ and $\deg q_1=3$ or $4$, the Chen iterated integrals can be expressed in terms of elliptic generalisations of MPLs evaluated at algebraic arguments.
\end{itemize}
In particular, we explain how we can algorithmically rewrite all Chen iterated integrals that meet these criteria in terms of MPLs and their elliptic analogues. This algorithm is well known, but we document it here because it will be an important tool to obtain compact analytic expressions for some master integrals that contribute to Bhabha scattering at two loops. 
Two comments are in order: First, in the presence of (multiple) square roots, it may be possible to change variables and rationalise (some of) the roots. In this way it may possible to reduce the problem to a situation covered by the criteria above, even though the original problem did not satisfy these conditions. Second, we stress that the aforementioned conditions are only sufficient, but by no means necessary, to rewrite Chen iterated integrals in terms of (elliptic) MPLs. In particular, there are several examples of Feynman integrals whose alphabets involve non-rationalisable square roots, but nevertheless, it was possible to express them in terms of ordinary MPLs, cf., ~e.g.,~refs.~\cite{Heller:2019gkq,Heller:2021gun,Kreer:2021sdt}. 

\subsection{Rational alphabets without square roots}
\label{sec:MPL_algo}
If the alphabet does not contain any square roots, or if all square roots can be rationalised, we can always express the Chen iterated integral in terms of MPLs evaluated at algebraic arguments. 
First, we can use the additivity of the logarithm to assume that all letters $R_i$ are irreducible polynomials. Next, we can use the homotopy-invariance of the integral to deform the contour $\gamma$ into a new contour $\gamma_0$ with the same end-points ${\bf x}_0$ and ${\bf x}_1$, without changing the value of the integral (except for picking up residues when we cross a pole). We choose the new contour as follows: We order the integration variables in some way, which for simplicity we assume to be the natural order $x_1,\ldots, x_p$. The contour $\gamma_0$ is then obtained as the concatenation of the straight-line segments $\gamma_r$, $1\le r \le p$, defined by ${\bf x} = \varphi^{(r)}(t)=(\varphi^{(r)}_{1}(t)\,\ldots,\varphi^{(r)}_{p}(t))$ with $0\le t\le 1$ and
\beq\label{eq:straight-line_parametrisation}
\varphi^{(r)}_{i}(t) = \left\{\begin{array}{ll}x_{1,i}\,,& \quad i < r\,,\\
\left(x_{1,r}-x_{0,r}\right)t + x_{0,r} \,, &\quad i=r\,,\\
x_{0,i}\,,&\quad i>r\,.
\end{array}\right.
\eeq
Next we can iteratively apply the path-composition formula for iterated integrals,
\beq
\int_{\gamma_1\gamma_2}\omega_{i_1}\ldots \omega_{i_n} = \sum_{k=0}^n\int_{\gamma_1}\omega_{i_1}\ldots \omega_{i_k}\int_{\gamma_2}\omega_{i_{k+1}}\ldots \omega_{i_n}\,,\qquad \omega_{i} = d\log R_i({\bf x})\,.
\eeq
The previous equation allows us to reduce the integral to a linear combination of products of Chen iterated integrals over the straight-line segments $\gamma_r$. On the line segment $\gamma_r$ the $d\log$-forms take a particularly simple form. Indeed, since all the letters $R_i$ are polynomials in ${\bf x}$, it is easy to see that $R_i(\varphi_r(t))$ is a polynomial in $t$. In the following we write 
\beq
R_i(\varphi_r(t)) = c_0\,(t-c_1)\ldots (t-c_d)\,,
\eeq
where $d$ denotes the degree of the polynomial $R_i(\varphi_r(t))$ and the quantities $c_j$, $0\le j\le d$ are algebraic functions of ${\bf x}_0$ and ${\bf x}_1$. The pull-back of the one-form $d\log R_i({\bf x})$ to the straight-line segment $\gamma_r$ then reads
\beq\label{eq:pull_back}
\gamma^{\ast}_rd\log R_i({\bf x}) =dt\,F_{r,i}(t)\,,
\eeq
with
\beq\label{eq:pull_back_function_MPLs}
F_{r,i}(t) = 
\left\{\begin{array}{ll}
0\,, & \textrm{ if } \partial_{x_{r}}R_i({\bf x}) = 0\,,\\
\sum_{j=1}^d\frac{1}{t-c_j}\,, & \textrm{ if } \partial_{x_{r}}R_i({\bf x}) \neq 0\,.
\end{array}\right.
\eeq
The previous equation makes it manifest that on each straight-line segment the Chen iterated integral evaluates to ordinary MPLs with algebraic arguments $c_j$, cf.~eq.~\eqref{eq:Mult_PolyLog_def}.

\subsection{Alphabets with a single elliptic square root} 
\label{sec:eMPL_algo}
We have already seen that a square root of a polynomial of degree three or four cannot be rationalised. More precisely, consider the set of points $(x,y)$ constrained by 
\beq
y^2=P_n(x)\,, 
\eeq
where $P_n(x)$ is a polynomial of degree $n$ in $x$. If $n=3$ or $4$, this equation defines an algebraic variety called an elliptic curve. 
It is thus not surprising that the case $N_{\textrm{sqrt}}=1$ and $q_1 = P_n$, with $n= 3$ or $4$, leads to generalisations of MPLs related to elliptic curves. We start by giving a lightning review of elliptic multiple polylogarithms (eMPLs), before we comment on the generalisation of the algorithm from section~\ref{sec:MPL_algo}. 
More details can be found in appendix~\ref{app:eMPLs}.

We focus here on the case $n=4$, and we assume that $P_4$ has the form $P_4(x)=(x-a_1)(x-a_2)(x-a_3)(x-a_4)$. 
Elliptic multiple polylogarithms (eMPLs) can then be defined as the iterated integrals~\cite{BrownLevin,Broedel:2014vla,Broedel:2017kkb,Broedel:2018qkq}
\beq\label{eq:cE4_def}
\cEf{n_1 & \ldots & n_k}{c_1 & \ldots& c_k}{x}{\vec{a}} = \int_0^xdt\,\Psi_{n_1}(c_1,t,\vec a)\,\cEf{n_2 & \ldots & n_k}{c_2 & \ldots& c_k}{t}{\vec a}\,,
\eeq
with $n_i\in\mathbb{Z}$ and $c_i\in{\mathbb{C}}\cup\{\infty\}$, and $\vec a=(a_1,a_2,a_3,a_4)$ is the vector of the four {branch points} $a_i$. Just like ordinary MPLs, eMPLs have at most logarithmic singularities, but no poles.
The number $n=\sum_{i=1}^k|n_i|$ is called the {weight} of the eMPL, and the number of integrations $k$ is its length. In the case where all the indices $A_i = \left(\begin{smallmatrix} n_i\\ c_i\end{smallmatrix}\right)$ are equal to $\left(\begin{smallmatrix} \pm 1\\ 0\end{smallmatrix}\right)$, the integral is divergent and requires a special treatment similar to the case $a_n=0$ for ordinary MPLs, cf.~eq.~\eqref{eq:GLog}. We refer to appendix~\ref{app:eMPLs} for details. Also note that we need to be careful about how we choose the branches of the square root $y=\sqrt{P_4(x)}$. We will come back to this point in section~\ref{sec:f14}.

There are infinitely many integration kernels $\Psi_{n}(c,x,\vec a)$ for given $(c,x,\vec a)$ in eq.~\eqref{eq:cE4_def}. In concrete applications only a finite number of these kernels appear. Here we only list the kernels with $|n|\le1$, which are of direct relevance to this paper.
For $n=0$, we have
\beq\label{eq:pure_psi0}
\Psi_0(0,x,\vec a) = \frac{c_4}{\omega_1\,y}\,,
\eeq
with $c_4 = \frac{1}{2}\sqrt{a_{13}a_{24}}$, $a_{ij} = a_i-a_j$. $\omega_1$ is one of the two {periods} associated to the elliptic curve defined by the equation $y^2 = P_4(x)$,
\beq
\omega_1 = 2\,\textrm{K}(\lambda)\,,\qquad \omega_2 = 2i\,\textrm{K}(1-\lambda)\,,\qquad \lambda = \frac{a_{14}\,a_{23}}{a_{13}\,a_{24}}\,,
\eeq
where K is the complete elliptic integral of the first kind
\beq
\textrm{K}(\lambda) = \int_0^1\frac{dt}{\sqrt{(1-t^2)(1-\lambda t^2)}}\,.
\eeq 
For $|n|=1$, we have (with $c\neq \infty$)
\begin{align}
\nonumber \Psi_1(c,x,\vec a) &\,= \frac{1}{x-c}\,, \\
\label{eq:pure_psi1}\Psi_{-1}(c,x,\vec a) &\,=  \frac{y_c}{y(x-c)} + Z_4(c,\vec a)\,\frac{c_4}{y}\,,\\
\nonumber\Psi_{1}(\infty,x,\vec a) &\,=  -Z_4(x,\vec a)\,\frac{c_4}{y}\,,\\
\nonumber\Psi_{-1}(\infty,x,\vec a) &\,=  \frac{x}{y}   -\frac{1}{y} \left[{a_1}+ 2c_4\, G_{\ast}(\vec a)\right]\,,
\end{align}
where we introduced the shorthand $y_c=\sqrt{P_4(c)}$. These functions have the property that the differential form $dx\,\Psi_{\pm1}(c,x,\vec a)$ has a simple pole at $x=c$, and no other poles (except for $dx\,\Psi_1(c,x,\vec a)$, which always has a pole at $x=\infty$). Note that the kernel $\Psi_1$ is identical to the kernel that defines ordinary MPLs (cf. eq.~\eqref{eq:Mult_PolyLog_def}), and so ordinary MPLs are a subset of eMPLs,
\beq\label{eq:cE4_to_G}
\cEf{1 & \ldots & 1}{c_1 & \ldots& c_k}{x}{\vec{a}} = G(c_1,\ldots,c_k;x)\,.
\eeq
The functions $Z_4(c,\vec a)$ and $G_{\ast}(\vec a)$ in eq.~\eqref{eq:pure_psi1} are in general transcendental functions. 
They can be expressed in terms of incomplete elliptic integrals of the first and second kind 
(see appendix~\ref{app:eMPLs}). Depending on the value of the argument $c$, in many applications it is possible to 
evaluate $Z_4(c,\vec a)$ and $G_{\ast}(\vec a)$ in the form $A+B\frac{i\pi}{\omega_1}$, 
where $A$ and $B$ are algebraic functions of $\vec a$ and $c$~\cite{Broedel:2018qkq}. 
In particular, in all cases relevant for this paper, $Z_4(c,\vec a)$ and $G_{\ast}(\vec a)$ will always be 
algebraic functions (see section~\ref{sec:bhabha_empls}).

Let us now assume that the alphabet is given by $d\log$-forms with arguments 
\beq
R_i(\bx) = r_{i,0}(\bx) + r_{i,1}(\bx)\,\sqrt{q_1(\bx)}\,,
\eeq
where $r_{i,0}(\bx)$ are polynomials in $\bx$ and $q_1(\bx)$ is a non-constant polynomial of degree at most four. We assume that the $r_{i,j}(\bx)$ do not have any common zero (otherwise we could factor out a term from this letter in the alphabet), and $q_1(\bx)$ does not have a double zero (otherwise we could factor it out of the square root). We allow $r_{i,1}(\bx)$ to be zero (in which case this letter does not depend on the square root), but $r_{i,0}(\bx)$ is assumed not to vanish. We can separate all the letters into even and odd parts:
\beq
d\log R_i(\bx) = \frac{1}{2}d\log R_i^+(\bx) - \frac{1}{2}d\log R_i^-(\bx)\,,
\eeq
with 
\beq\bsp
R_i^+(\bx) &\,=  r_{i,0}(\bx)^2 - r_{i,1}(\bx)^2{q_1(\bx)}\,,\\
R_i^-(\bx) &\,= \frac{ r_{i,0}(\bx) - r_{i,1}(\bx)\,\sqrt{q_1(\bx)}}{ r_{i,0}(\bx) + r_{i,1}(\bx)\,\sqrt{q_1(\bx)}}\,.
\esp\eeq

Next we deform the integration path $\gamma$ to the sequence of line segments $\gamma_r$ defined in eq.~\eqref{eq:straight-line_parametrisation}. 
Even letters give rise to integration kernels of the form $\frac{dt}{t-c_j}$ and can be dealt with exactly as in section~\ref{sec:MPL_algo}. We will not discuss them any further. For the odd letters, note that on the straight-line segment $\gamma_r$, $q_1(\varphi_r(t))$ is a polynomial of degree $n$ in $t$, with $n\le 4$. We write:
\beq
q_1(\varphi_r(t)) = \alpha_r\,P^{(r)}(t) = \alpha_r\,(t-a_{r,1})\cdots (t-a_{r,n})\,,\qquad \vec a_{r} = (a_{r,1},\ldots, a_{r,n})\,.
\eeq
 If $n\le 2$, then the square root $\sqrt{P^{(r)}(t)}$ can be rationalised, and all letters are rational on the line segment $\gamma_r$, and we do not need to discuss this case anymore. If the degree of $P^{(r)}(t)$ is three or four, then we cannot rationalise the square root on $\gamma_r$. Instead, we obtain
\beq
\gamma_r^*d\log R_i^-(\varphi_r(t)) = dt\,F^-_{r,i}(t)\,,
\eeq
where the parity of $R_i^-$ implies that $F^-_{r,i}(t)$ is of the form
\beq
F^-_{r,i}(t) = \frac{1}{\sqrt{P^{(r)}(t)}}\,S_{r,i}(t)\,,
\eeq
where $S_{r,i}(t)$ is a rational function in $t$. At this point we make an important observation: By construction, the differential form $d\log R_i^-(\varphi_r(t))$ has only logarithmic singularities. This implies that $S_{r,i}(t)$ has poles of order at most one (possibly including a simple pole at infinity). As a consequence, $S_{r,i}(t)$ may be written as a linear combination of the functions $\Psi_{-1}(c_j,t,\vec a_r)$ and $\Psi_{0}(c_j,t,\vec a_r)$. Hence, we can evaluate the integral on $\gamma_r$ in terms of eMPLs for the elliptic curve associated to $P^{(r)}(t)$. Note that, a priori, we may obtain eMPLs with different values of $\vec a_r$ for each segment $\gamma_r$.

\section{Integration of the differential equations in terms of MPLs}
\label{sec:can_diff_eqs_part2}
After the general remarks of the previous section, we go back to the explicit solution for the system in eq.~\eqref{eq:DEs-ep2}.
The standard way to rationalize the first two square roots in eq.~\eqref{eq:sqrts} is to turn to
the dimensionless Landau variables $x$ and $y$ related to the Mandelstam invariants by
\beq
\label{eq:xyvar}
\frac{-s}{m^2} = \frac{(1-x)^2}{x}\textrm{~~and~~}\frac{-t}{m^2} = \frac{(1-y)^2}{y}\,.
\eeq
In terms of these variables the Euclidean region $s,t<0$ corresponds to $0\le x,y\le 1$ 
(using the symmetry of eq. \eqref{eq:xyvar} under $(x,y)\to(1/x,1/y)$). 

There exist different strategies to solve the differential equation in eq.~\eqref{eq:DEs-ep2}. The first method consists in evaluating numerically the $\eps$-expansion of the path-ordered exponential in eq.~\eqref{eq:g_path_sol}. 
This can be achieved by application of the Frobenius method to look for power-series solutions of ordinary differential 
equations in the vicinity of regular singular points. 
The Frobenius method has been successfully used in various one-dimensional problems in the past~\cite{Pozzorini:2005ff,Aglietti:2007as,Caffo:2008aw,Bonciani:2018uvv},
and more recently a generalisation of this method has been proposed in ref.~\cite{Francesco:2019yqt} 
to deal with complicated multidimensional problems, see for example~\cite{Bonciani:2019jyb,Abreu:2020jxa}.
This strategy has also been implemented into the public code 
code {\tt DiffExp}~\cite{Hidding:2020ytt}.\footnote{A Mathematica notebook is provided as ancillary material with the arXiv submission.}  
The input data for this code are the matrices that define the differential equations, and the boundary conditions in some limit, e.g., at some point $(x_0,y_0)$. The code then uses this input to evolve the solution numerically from the point $(x_0,y_0)$ to some generic point $(x,y)$. We fix the boundary conditions in the limit $s,t\to 0$, which corresponds to $(x_0,y_0)=(1,1)$ in terms of the Landau variables in eq.~\eqref{eq:xyvar}. Using the expansion by regions strategy 
implemented
in the public code {\tt asy.m}~\cite{Pak:2010pt,Jantzen:2012mw} (which is now included in 
the code {\tt FIESTA} \cite{Smirnov:2015mct}) we obtain the following leading order asymptotic behaviour in this limit
\begin{align}
\nonumber f_{1}&\,\sim 1+\frac{\pi ^2\epsilon ^2}{6} -\frac{2 \zeta_3 \epsilon ^3}{3}+\frac{7 \pi ^4 \epsilon ^4}{360} + \ord(\eps^5)\,, \\   
\nonumber f_{6}&\,\sim-\frac{1}{4} -\frac{5 \pi^2 \epsilon ^2}{24} -\frac{11 \zeta_3 \epsilon ^3}{6}-\frac{101}{480} \pi ^4 \epsilon ^4+ \ord(\eps^5)\,, \\   
\label{eq:bc_st0} f_{9}&\,\sim 
 -\frac{\pi ^2 \epsilon ^2}{12} + \frac{1}{4} \epsilon ^3 \left(2
   \pi ^2 \log 2-7 \zeta_3\right)   
   \\
 \nonumber&\, +\frac{1}{180} \epsilon ^4 \left(13 \pi
   ^4-90 \log ^42-180 \pi ^2 \log ^22-2160 \text{Li}_4\left(\frac{1}{2}\right)\right) + \ord(\eps^5) \,, \\ 
 \nonumber f_{18}&\,\sim 
\frac{1}{2} \epsilon ^3 \left(2 \pi ^2
   \log 2-3 \zeta_3\right)
+\frac{1}{20} \epsilon ^4 \left(7 \pi ^4-20
   \log ^42-40 \pi ^2 \log ^22-480 \text{Li}_4\left(\frac{1}{2}\right)\right)+ \ord(\eps^5)
   \,, \\ 
\nonumber f_{19}&\,\sim (-s)^{-\epsilon } \left(-1+\frac{8 \zeta_3 \epsilon ^3}{3}+\frac{\pi ^4 \epsilon
   ^4}{30}\right)+ \ord(\eps^5) \,, \\ 
\nonumber f_{22}&\,\sim (-s)^{-\epsilon } \left(-\frac{1}{2}+\frac{4 \zeta_3 \epsilon ^3}{3}
+\frac{\pi ^4 \epsilon^4}{60}\right)+(-s)^{-2 \epsilon } \left(\frac{1}{4}-\frac{\pi ^2 \epsilon^2}{24}-\frac{14 \zeta_3
   \epsilon ^3}{3}-\frac{67}{480} \pi ^4 \epsilon ^4\right)+ \ord(\eps^5) \,, \\ 
\nonumber f_{23}&\,\sim (-s)^{-2 \epsilon } \left(\pi ^2 \epsilon ^2+2 \pi^2 \epsilon ^3 \log 2
+2 \pi ^2 \epsilon ^4 \left(\pi ^2+\log ^22\right)\right)+ \ord(\eps^5) \,, \\ 
\nonumber f_{25}&\,\sim (-s)^{-\epsilon } \left(-\pi ^2 \epsilon ^2-2 \pi ^2 \epsilon ^3 \log 2
-\frac{1}{2} \pi ^2 \epsilon ^4 \left(\pi ^2+4 \log^22\right)\right)+ \ord(\eps^5)\,,
\end{align}
and $f_i\sim 0$, i.e., $f_i=\ord(s,t)$, for all the other elements. 

Let us emphasise that, in order to profit maximally from the automated code {\tt DiffExp}, 
it is crucial that the input data are in an optimal form.
This includes providing differential equations in canonical form and fixing the boundary conditions in a 
simple point ($s=t=0$, i.e., $x=y=1$). 
With our input, the code works very well and provides the possibility to obtain high-precision 
numerical results (100 digits accuracy and more), both in the Euclidean and the physical regions. 
In other words, {\tt DiffExp} allows us to obtain high-precision numerical results for all master integrals and for all values of the input parameters. The code runs fast enough to allow its usage for practical applications. 
This is then of course sufficient for all phenomenological applications one has in mind.
However, both from a formal and from a practical point of view, it may still 
be desirable in some situations to have full-fledged analytic representations of the master integrals in terms of functions 
that are well studied in the literature, e.g., MPLs. 
In the remainder of this section we explain how we can obtain analytic results for all master integrals but $f_{14}$
 in terms of MPLs evaluated at 
algebraic arguments. The reason why $f_{14}$ is different will be discussed 
in section~\ref{sec:f14}. Here it suffices to say that the square root $r_u$ only enters the differential equation for $f_{14}$. 
Since our strategy of solving the differential equations in terms of MPLs will be closely based on the ideas from section~\ref{sec:MPL_algo}, it is not suprising that an additional square root implies that the sufficient condition of section~\ref{sec:MPL_algo} is not satisfied. We will discuss the case of $f_{14}$ in detail in section~\ref{sec:f14}, 
and we focus for the rest of this section only on the other master integrals.
\newline

In order to solve the master integrals in terms of MPLs, we start by observing that 
the square root $r_{st}$ does not appear when solving the differential equations up to weight 3 for all elements but $f_{37}$, and
at weight 4 for all elements but $f_{i}, i\in\{35, 36, 37, 38, 39, 41, 43\}$. The equations can be solved by integrating first in $x$, 
resulting in MPLs of the form $G(\vec c;x)$ with $c_i\in\{0, -1, 1, -y, -1/y\}$. This allows us to fix the solution up to a function of $y$,
which can be determined by substituting the results of the $x$-integration into the differential equations in $y$,  checking that the variable $x$ disappears in them,
and finally solving them in terms of MPLs of the form $G(\vec c;y)$ with $c_i\in\{0, -1, 1\}$, i.e., harmonic polylogarithms~\cite{Remiddi:1999ew}.
At this point the solution is fixed up to a set of undetermined integration constants, which we fix using the boundary conditions in eq.~\eqref{eq:bc_st0}.

To evaluate $f_{37}$ at weights 3 and 4 and $f_{i}$ with $i\in\{35, 36, 38, 39, 41, 43\}$ at weight 4, we have to deal with
the square root $r_{st}$. It can be rationalized by the following further change of variables
\beq\label{eq:fromxtow}
x = 2\,\frac{(1-w) \left(y^2-y+1\right)^2-2
   y^2}{\left(1-w^2\right) \left(y^2-y+1\right)^2} \,.
\eeq
The resulting system of differential equations for 42 elements can be solved, first
in $w$ and then in $y$. Solving in $w$ gives a linear combination of MPLs $G(\ldots;w)$ 
with the letters the alphabet $\{l^w_1,\ldots,l^w_{15}\}$, where $l^w_i, i=1,\ldots,11$, are taken from the set
\beq\bsp\label{eq:w-alphabet}
&\,\left\{1,-1,\frac{y^4-2 y^3+y^2-2
   y+1}{\left(y^2-y+1\right)^2},\frac{y^2+y+1}{y^2-y+1},\frac{y^2-3
   y+1}{y^2-y+1},\right.\\   
 &\,  -\frac{y^2+2 \sqrt{y^2+1} y-2
   \sqrt{y^2+1}-y+1}{y^2-y+1},-\frac{y^2-2 \sqrt{y^2+1} y+2
   \sqrt{y^2+1}-y+1}{y^2-y+1},\\
&\,\left.\frac{y^2-y-1}{y^2-y+1},-\frac{y^2+y-1}{y^2-y+1},
-\frac{2 y^3-y^2+y-1}{y^2-y+1},\frac{y^3-y^2+y-2}{y \left(y^2-y+1\right)}\right\} \,,
\esp\eeq
and $l^w_i, i=12,\ldots,15$ are the roots of the polynomial $\cQ(y,w)$ defined by


\begin{align}
\cQ(y,w) &= w^4 \left(y^2-y+1\right)^4 y+2 w^3 \left(y^2-4 y+1\right) \left(y^2-y+1\right)^4  \\
&-2 w^2 \left(y^2-y+1\right)^2
   \left(y^6-7 y^5+12 y^4-11 y^3+12 y^2-7 y+1\right) \nonumber \\
&-2 w \left(y^2-y+1\right)^2 \left(y^6-2 y^5+4 y^4-12 y^3+4 y^2-2 y+1\right) \nonumber \\
&+2 y^{10}-11 y^9+30 y^8-62 y^7+90 y^6-89 y^5+90 y^4-62 y^3+30 y^2-11 y+2\,.\nonumber
\end{align}
Solving the differential equation in $y$ (where the dependence on $w$ drops out) gives a linear combination of MPLs 
$G(\ldots;y)$ with the letters from the set $\{l^y_1,\ldots,23\}$, with
\beq\bsp\label{eq:y-alphabet}
&\,l^y_{1} =  0\,, \,l^y_{6} =  \frac{1}{2}
   \left(3-\sqrt{5}\right)\,, \,l^y_{7} =  \frac{1}{2} \left(3+\sqrt{5}\right)
 \,,  \,l^y_{8} =  e^{\frac{i \pi }{3}}\,, \,l^y_{9} = 
   e^{-\frac{i \pi }{3}} \,,\,l^y_{10} =  -e^{\frac{i \pi }{3}}\,, \\
& \,l^y_{11} = 
   -e^{-\frac{i \pi }{3}}\,,\,l^y_{12} =  1\,,\,l^y_{13} =  -1\,, \,l^y_{24} = 
   \frac{1}{2} \left(1-\sqrt{5}\right)\,, \,l^y_{25} =  \frac{1}{2}
   \left(1+\sqrt{5}\right)\,,\\
&\,l^y_{26} =  \frac{1}{2}
   \left(-1-\sqrt{5}\right)\,, \,l^y_{27} =  \frac{1}{2}
   \left(\sqrt{5}-1\right) \,.
\esp\eeq
Moreover,   
$l^y_i, i=2,\ldots,5$ are the roots of $1 - 2 y + y^2 - 2 y^3 + y^4$; 
$l^y_i, i=14,\ldots,17$ are the roots of $3 - 6 y + 5 y^2 - 6 y^3 + 3 y^4$;
$l^y_i, i=18,\ldots,20$ are the roots of $-2 + y - y^2 + y^3$ and
$l^y_i, i=21,\ldots,23$ are the roots of $-1 + y - y^2 + 2 y^3$.
 
Following this procedure, we obtain complete analytical results for $f_{37}$ at weights 3 and 4, and for the elements $f_{i}$ with $i\in\{35, 36, 38, 39, 41, 43\}$ at weight 4,
in terms of MPLs in $w$ or $y$ with the letters defined above.  We have checked our analytic results for the master integrals against {\tt FIESTA} \cite{Smirnov:2015mct} using also the forthcoming 
new release \cite{Smirnov:fiesta2021}.
It turns out,
however, (as it is easy to imagine from the dimension of the alphabet) that our results at
weight 4 for these elements are rather complicated due to a very intricate branch-cut structure.
In particular, evaluating them in this form 
at a phase-space point with
{\tt GiNaC}~\cite{Bauer:2000cp,Vollinga:2004sn} meets problems connected both with timing and stability, 
so that our results for the 
complicated elements become impractical, and for phenomenological applications the numerical solution obtained
using {\tt DiffExp} might be preferable. 
For the same reason, we also recommend to turn to DiffExp also for the contribution of element 37 of weight 3.
Of course, this does not imply that a better analytical representation in terms of MPLs does not exist,
whose numerical evaluation could be much faster than automated numerical codes. 
Our analytic results obtained by direct integration of the differential equations 
could be used as a starting point to look for such a better
analytic representation, if required.




\section{Compact analytic results in terms of eMPLs}
\label{sec:f14}

\subsection{An analytic result for $f_{14}$ in terms of eMPLs}
\label{sec:bhabha_empls}

As mentioned in the previous section, we could solve the differential equations for all master integrals analytically in terms of MPLs with algebraic arguments, except for $f_{14}$. The reason why we cannot easily apply the techniques from the previous section to $f_{14}$ can be seen from the form of the differential equation itself. If we write $f_{14}(x,y) = \eps^4\,\bar{f}(x,y)$, then the differential equation in $x$ takes the form:
\begin{align}
\nonumber&\,\frac{\partial}{\partial x}\bar{f}(x,y)=\frac{1}{(x-1) x \sqrt{\Delta(x,y)}} 
\Bigg\{ -2(x+1)  x \left(y^2-1\right)G_0(y) 
   G_{0,0}(x)\\
\label{eq:dxfbar}&\,\phantom{+}(x-1) G_0(x) \Big[ 2 \left(3 x^2 y+x (y-1)^2+y\right) G_{0,0}(y)+\pi ^2
   \left(x^2-1\right) y \Big] \\
\nonumber   &\, -(x+1)(x-1)^2 y \Big[ 2 G_0(y)  \left(G_{-1/y,0}(x)-G_{-y,0}(x)\right)+2G_{-1/y,0,0}(x)+2G_{-y,0,0}(x)\\
 \nonumber  &\,+2 G_{0,0,0}(x)+4
   G_{1,0,0}(x) -4G_{0,0,0}(y)+4 G_{1,0,0}(y)+2\zeta_3+ \left(2
   G_{0,0}(y)+\pi ^2\right) G_{-1/y}(x)
 \\
\nonumber &\,  + \left(2
  G_{0,0}(y)+\pi ^2\right) G_{-y}(x) \Big] \Bigg\} \,,
\end{align}
with
\beq\label{eq:Delta_def}
\Delta(x,y) = (x+y) (x y+1) \left(x^2 y+xy^2-4 x y+x+y\right)\,.
\eeq
 Due to the symmetry $f_{14}(x,y) = f_{14}(y,x)$, the differential equation with respect to $y$ can easily be obtained by exchanging the roles of $x$ and $y$ in eq.~\eqref{eq:dxfbar}. The square root in eq.~\eqref{eq:dxfbar} is identical to the square root that has appeared in ref.~\cite{Henn:2013woa} in the computation of the first planar two-loop family for Bhabha scattering. The polynomial under the square root has degree four in both $x$ and $y$, so it cannot be rationalised in one variable only. However, this is not sufficient to exclude that one cannot rationalise it by performing a change of variable involving simultaneously $x$ and $y$. In ref.~\cite{Festi:2018qip} it was shown that the algebraic variety defined by $\xi^2=\Delta(x,y)$ is a K3 surface. As a consequence, it cannot be rationalised by any rational change of variables, and the strategy of rationalising all square roots and evaluating all iterated integrals in terms of MPLs using ideas from section~\ref{sec:MPL_algo} cannot be applied. 
We can of course also obtain numerical results for $f_{14}$ by solving the canonical system with {\tt DiffExp} (as described in the previous section), but it would of course be interesting to obtain analytic results also for $f_{14}$.

We observe that the structure of the square root matches precisely the criteria of section~\ref{sec:eMPL_algo}. Indeed, the polynomial under the square root in eq.~\eqref{eq:dxfbar} has degree four in both $x$ and $y$. So, if we keep one variable fixed, the square root defines an elliptic curve in the other. This reflects the fact that the K3 surface defined by the square root is elliptically fibered~\cite{Festi:2018qip}. As a consequence, we can choose the integration path in eq.~\eqref{eq:straight-line_parametrisation} and solve the differential equation in terms of eMPLs. Before we do this, we have to make an important comment. While the strategy of section~\ref{sec:MPL_algo} to solve eq.~\eqref{eq:dxfbar} does not apply, it does not exclude that a solution in terms of MPLs evaluated at algebraic arguments exists. On the contrary, we have found that 
it is possible to evaluate $f_{14}$ from its Feynman parameter representation using direct integration techniques (cf., e.g., refs.~\cite{Brown:2009qja,Brown:2008um,Anastasiou:2013srw,Ablinger:2014yaa,Bogner:2014mha,Panzer:2014caa,Duhr:2019tlz}).\footnote{We are grateful to Erik Panzer for providing us with a change of variables that renders the Feynman parameter integral linearly reducible.} The resulting expression, however, is extremely lengthy and involves MPLs evaluated at complicated algebraic arguments. We have not been able to confirm the final expression numerically, as its size and the complexity of the branch cuts renders the evaluation of the MPLs extremely challenging. Our representation obtained from direct integration is therefore not useful for practical purposes. As we will show now, the representation in terms of eMPLs is very compact.

Let us now explain how we can solve the system of differential equations for $\bar{f}$ in terms of MPLs and eMPLs. 
We first of all introduce new variables $(\bar{x},\bar{y}) = (1-x,1-y)$, and  
use {\tt PolyLogTools}~\cite{Duhr:2019tlz} to express all MPLs of the form $G(\ldots;{x})$ or $G(\ldots;{y})$ in eq.~\eqref{eq:dxfbar} in terms of $G(\ldots;\bar{x})$ and $G(\ldots;\bar{y})$ respectively. For example, we find:
\beq\bsp
G_{-y,0,0}(x) &\,= G_{2-\bar{y},1,1}(\bar{x})-\frac{\pi ^2}{12}\,  G_1(\bar{y})-G_{1,1,1}(\bar{y})+G_{1,1,2}(\bar{y})+\log2\, G_{1,1}(\bar{y})+\frac{3 }{4}\,\zeta_3\,.
\esp\eeq

We know from eq.~\eqref{eq:bc_st0} that $f_{14}$ must vanish for $s=t=0$, and so 
$\bar{f}(\bar{x}=0,\bar{y}=0)=0$.  
Our strategy is then as follows. We first use the differential equation in $\bar{y}$ to evolve $\bar{f}$ from 
$(\bar{x},\bar{y})=(0,0)$ to 
$(\bar{x},\bar{y})=(0,\bar{y}_0)$. We then use the differential equation in $\bar{x}$ to evolve from  
$(\bar{x},\bar{y})=(0,\bar{y}_0)$ to the generic point $(\bar{x},\bar{y})=(\bar{x}_0,\bar{y}_0)$.
In the following we assume $0<\bar{x}_0,\bar{y}_0<1$ for concreteness (which corresponds to he Euclidean region $s,t<0$).
On the line $\bar{x}=0$ we find
\beq
\Delta(x=1,y) = (1-y^2)^2 = \bar{y}^2(2-\bar{y})^2\,.
\eeq
Hence, the square root disappears in the limit $\bar{x}\to0$, and we can solve the differential equation on the line ${\bar{x}}$ in terms of MPLs. We find (relabelling $\bar{y}_0$ as $\bar{y}$):
\beq\bsp
\bar{f}(\bar{x}=0,\bar{y}) &\,= \left(2 \pi ^2 \log 2-3 \zeta_3\right)\, G_1(\bar{y})-\pi ^2\, G_{1,1}(\bar{y})+2 \pi ^2\, G_{1,2}(\bar{y})\\
&\,+4\, G_{1,0,1,1}(\bar{y})-4 \,G_{1,1,1,1}(\bar{y})+4\, G_{1,2,1,1}(\bar{y})\,.
\esp\eeq

Next, let us discuss the solution on the line $\bar{y}=\bar{y}_0$. Clearly, $\Delta(x,y)$ is not a perfect square for $x\neq 0$. Looking at the constraint $\xi^2=\Delta(x,y_0)$ as a function of $x$ with $y_0$ fixed, we see that it defines an elliptic curve. The branch points, i.e., the zeroes in $\bar{x}$ of $\Delta(1-\bar{x},1-\bar{y}_0)$, are
\beq\label{eq:f14_bps}
\vec a(\bar{y}_0) = \left(2-\bar{y}_0,\frac{\bar{y}_0 \left(\bar{y}_0-\sqrt{\bar{y}_0^2+4 \bar{y}_0-4}\right)}{2 \left(1-\bar{y}_0\right)},\frac{\bar{y}_0\left(\bar{y}_0+\sqrt{\bar{y}_0^2+4 \bar{y}_0-4}\right)}{2 \left(1-\bar{y}_0\right)},\frac{2-\bar{y}_0}{1-\bar{y}_0}\right)\,.
\eeq
We can use eq.~\eqref{eq:cE4_to_G} to interpret every MPL of the form $G(\vec b(\bar{y});\bar{x})$ as an eMPL of the form $\cEf{\vec n }{ \vec c(\bar{y})}{\bar{x}}{\vec a(\bar{y})}$. 
Moreover, we can express all the algebraic coefficients multiplying the MPLs in the differential equation in terms of the functions $\Psi_{\pm1}(c,\bar{x},\vec{a}(\bar{y}))$. For example, we find
\beq\bsp
\Psi_{-1}(\infty,\bar{x},\vec a(\bar{y})) &\,= \frac{\bar{x}}{\xi}-\frac{(y+1)^2}{4y \xi}\,,\\
\Psi_{-1}(0,\bar{x},\vec a(\bar{y})) &\,= \frac{y^2-1}{ \bar{x} y\xi}-\frac{y^2-1}{2y\xi}\,,\\
\Psi_{-1}(1,\bar{x},\vec a(\bar{y})) &\,= \frac{(y-1)^2}{4y \xi }-\frac{1}{(\bar{x}-1)\xi}\,,
\esp\eeq
with $\xi = \sqrt{\Delta(1-\bar{x},1-\bar{y})}$. In the process, we discover the relations:
\beq\bsp
Z_4(0,\vec a(\bar{y}))&\,= \frac{\left(2-\bar{y}\right) \bar{y}}{2 c_4 \left(1-\bar{y}\right)} = \frac{1-y^2}{2yc_4}\,,\\
Z_4(1,\vec a(\bar{y}))&\,= \frac{\bar{y}^2}{4 c_4 \left(1-\bar{y}\right)} = \frac{(1-y)^2}{4yc_4}\,,\\
G_{\ast}(\vec a(\bar{y})) &\,= \frac{\left(2-\bar{y}\right) \left(3 \bar{y}-2\right)}{8 c_4 \
\left(1-\bar{y}\right)} = \frac{(1-y)(1-3y)}{8yc_4}\,.
\esp\eeq
At this point we need to make an important comment about how we choose the branches of the square root $\xi=\sqrt{1-\bar{x},1-\bar{y})}$. From eq.~\eqref{eq:f14_bps} we can see that for $0<y<3-2\sqrt{2}$, the four branch points are real and ordered according to $a_1(y)<a_2(y)<a_3(y)<a_4(y)$. The branches of the square root are chosen according to
\beq
\xi=\sqrt{\Delta(x,y)} = \sqrt{|\Delta(x,y)|}\times\left\{\begin{array}{ll}
-1\,,& x<a_1(y)\textrm{ or }x\ge a_4(y)\,,\\
-i\,,& a_1(y)\le x<a_2(y)\,,\\
\phantom{-}1\,,&  a_2(y)\le x<a_3(y)\,,\\
\phantom{-}i\,,&  a_3(y)\le x<a_4(y)\,.
\end{array}\right.
\eeq
For $3-2\sqrt{3}<y<1$, we have $a_1(\bar{y}_0),a_4(\bar{y}_0)>0$ and $a_2(\bar{y}_0)^*=a_3(\bar{y}_0)$, with $\textrm{Im }a_3(\bar{y}_0)>0$. The branches of the square are then chosen as:
\beq
\xi=\sqrt{\Delta(x,y)} = \sqrt{|\Delta(x,y)|}\times\left\{\begin{array}{ll}
-1\,,& x<a_1(y)\textrm{ or }x\ge a_4(y)\,,\\
-i\,,& a_1(y)\le x<a_4(y)\,.
\end{array}\right.
\eeq
We stress that part of this choice is purely conventional, and compensated by the leading singularity $1/\sqrt{\Delta(x,y)}$ when passing from the master integral $g_{14}$ to its canonical analogue $f_{14}$.

The resulting differential equation can be solved by quadrature, and all integrals over $\bar{x}$ can be performed in terms of eMPLs using eq.~\eqref{eq:cE4_def}.
The final result reads (we relabel again $(x_0,y_0)$ as $(x,y)$):
\begin{align}
\nonumber \bar{f}(x,y) &\,= 2\cEf{-1 & 1 & 1 & 1}{\infty & \frac{1}{y}+1 & 1 & 1}{\bar{x}}{\vec{a}}+2\cEf{-1 & 1 & 1 & 1}{\infty & y+1 & 1 & 1}{\bar{x}}{\vec{a}}+2\cEf{-1 & 1 & 1 & 1}{1 & \frac{1}{y}+1 & 1 & 1}{\bar{x}}{\vec{a}}\\
\nonumber&\,+2\cEf{-1 & 1 & 1 & 1}{1 & y+1 & 1 & 1}{\bar{x}}{\vec{a}}+4\cEf{-1 & 1 & 1 & 1}{\infty & 0 & 1 & 1}{\bar{x}}{\vec{a}}-4\cEf{-1 & 1 & 1 & 1}{\infty & 1 & 1 & 1}{\bar{x}}{\vec{a}}\\
\nonumber&\,+4\cEf{-1 & 1 & 1 & 1}{1 & 0 & 1 & 1}{\bar{x}}{\vec{a}}-4\cEf{-1 & 1 & 1 & 1}{1 & 1 & 1 & 1}{\bar{x}}{\vec{a}}-\left(3\log^2y+\pi^2\right)\cEf{-1 & 1}{\infty & 1}{\bar{x}}{\vec{a}}\\
\label{eq:fbar_eMPL}
&\,+\left(\log^2y+\pi^2\right)\Big[\cEf{-1 & 1}{\infty & \frac{1}{y}+1}{\bar{x}}{\vec{a}}+\cEf{-1 & 1}{\infty & y+1}{\bar{x}}{\vec{a}}+\cEf{-1 & 1}{1 & \frac{1}{y}+1}{\bar{x}}{\vec{a}}\\
\nonumber&\,\quad+\cEf{-1 & 1}{1 & y+1}{\bar{x}}{\vec{a}}\Big]+\left(\log^2y-\pi^2\right)\cEf{-1 & 1}{1 & 1}{\bar{x}}{\vec{a}}+2\log y\Big[\cEf{-1 & 1 & 1}{\infty & \frac{1}{y}+1 & 1}{\bar{x}}{\vec{a}}\\
\nonumber&\,\quad-\cEf{-1 & 1 & 1}{\infty & y+1 & 1}{\bar{x}}{\vec{a}}+2\cEf{-1 & 1 & 1}{0 & 1 & 1}{\bar{x}}{\vec{a}}+\cEf{-1 & 1 & 1}{1 & \frac{1}{y}+1 & 1}{\bar{x}}{\vec{a}}-\cEf{-1 & 1 & 1}{1 & y+1 & 1}{\bar{x}}{\vec{a}}\Big]\\
\nonumber&\,+\Big[-4\text{Li}_3(-y)-4\text{Li}_3(y)+4\text{Li}_2(-y)\log y +4\text{Li}_2(y)\log y -\frac{2}{3}\log^3y\\
\nonumber&\,\quad+2\log(1-y)\log^2y+2\log(y+1)\log^2y-\pi^2\log y+2\pi^2\log(y+1)\\
\nonumber&\,\quad-2\zeta_3\Big]\Big[\cEf{-1}{\infty}{\bar{x}}{\vec{a}}+\cEf{-1}{1}{\bar{x}}{\vec{a}}\Big]\\
\nonumber&\,-12\text{Li}_4(-y)-12\text{Li}_4(y)-2\text{Li}_2(y)\log^2y-2\text{Li}_2(-y)\left(\log^2y+\pi^2\right)\\
\nonumber&\,+8\text{Li}_3(-y)\log y+8\text{Li}_3(y)\log y -2\zeta_3\log y-\frac{1}{6}\log^4y-\frac{1}{2}\pi^2\log^2y-\frac{3\pi^4}{20}\,,
\end{align}
where we introduced the shorthand $\vec a \equiv \vec a(\bar{y})$, and we have replaced all MPLs by classical polylogarithms:
\beq
\textrm{Li}_n(z) = -G(\underbrace{0,\ldots,0}_{n-1},1;z)\,.
\eeq
We have checked numerically that our final result for $\bar{f}(x,y)$ is correct by comparing against {\tt Fiesta}. The eMPLs in eq.~\eqref{eq:fbar_eMPL} were evaluated numerically using an in-house code (a public numerical implementation of eMPLs into {\tt GiNaC} exists~\cite{Walden:2020odh}). Note that $\bar{f}(x,y)$ is a pure function of uniform weight four~\cite{Broedel:2018qkq}. We find it interesting that such a compact analytic expression in terms of eMPLs exists, while our MPL expression obtained by naive direct integration is prohibitively large.

\subsection{A compact analytic result for the first planar family}
\label{sec:old_empls}

\begin{figure}[!t]
\begin{center}
\includegraphics{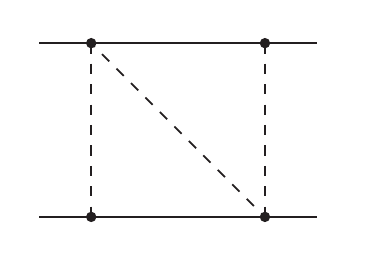}
\caption{The graph associated with the integral $B(s,t,m^2)$ in eq.~\eqref{eq:B_def}.
}
\label{fig:bhabha_empl_int}
\end{center}
\end{figure}

Let us conclude this paper by applying the techniques from section~\ref{sec:bhabha_empls} to the planar family in fig.~\ref{plaBha} (a). In ref.~\cite{Henn:2013woa} it was shown that all integrals in this family can be expressed in terms of 23 master integrals, and all but one master integral were evaluated in terms of MPLs. The remaining master integral that could not be expressed in terms of MPLs in ref.~\cite{Henn:2013woa} is (see fig.~\ref{fig:bhabha_empl_int}):
\begin{align}
\label{eq:B_def}
B(s,t,m^2) &\,=  \frac{e^{2\gamma_E\eps}}{\pi^D}\int\frac{d^Dk_1\,d^Dk_2}{[(k_1+p_1+p_2)^2-m^2][k_2^2-m^2](k_1+p_1)^2(k_1-k_2)^2(k_2-p_3)^2}\,.
\end{align}
This integral was evaluated in terms of a rather lengthy combination of MPLs with algebraic arguments in ref.~\cite{Heller:2019gkq}. Here we show that, by using the same strategy as in section~\ref{sec:bhabha_empls}, we can obtain a very compact representation for $B(s,t,m^2)$ in terms of the same type of eMPLs as in eq.~\eqref{eq:fbar_eMPL}.

The integral in eq.~\eqref{eq:B_def} is finite in four dimensions. Let us define $\widetilde{B}(x,y)$ by
\beq
B(s,t,m^2) = \frac{1}{4\sqrt{(s+t)(s+t-4m^2)}}\,\widetilde{B}(x,y) + \ord(\eps)\,,
 \eeq
 where the Landau variables $(x,y)$ have been defined in eq.~\eqref{eq:xyvar}. 
Our starting point is the differential equation satisfied by $\widetilde{B}(x,y)$~\cite{Henn:2013woa}:
\begin{align}\label{eq:dBtilde}
 d\widetilde{B}(x,y) &\,= g_1\,d\log\frac{1-Q(x,y)}{1+Q(x,y)} +g_2\,d\log\frac{(1+x)+(1-x)Q(x,y)}{(1+x)-(1-x)Q(x,y)} \\
 &\,+g_3\,d\log\frac{(1+y)+(1-y)Q(x,y)}{(1+y)-(1-y)Q(x,y)}\,.
 \end{align}
The functions $g_i$ appearing in the right-hand side of eq.~\eqref{eq:dBtilde} are pure combinations of MPLs of weight three,

\beq\bsp
g_1&\,=-16\,G_0(y)\,G_{0,0}(x)+32\,G_1(y)\,G_{0,0}(x)+8\,G_{0,0,0}(y)+16\,G_{0,0,1}(y)\\
&\,-32\,G_{0,-1,0}(x)+8\,G_{0,0,0}(x)+16\,G_{1,0,0}(x)+\frac{8}{3}\pi^2\,G_{0}(y)-\frac{16}{3}\pi^2\,G_1(y)\\
&\,-\frac{8}{3}\pi^2\,G_{0}(x)-32\,\zeta_3\,,\\
g_2&\,=-16\,G_{0,0,0}(x)-\frac{8}{3}\pi^2\,G_{0}(x)\,,\\
g_3&\,=8\,G_0(x)\,G_{-\frac{1}{x},0}(y)+16\,G_0(x)\,G_{-\frac{1}{x},1}(y)-8\,G_0(x)\,G_{-x,0}(y)-\frac{4}{3}\pi^2\,G_{0}(x)\\
&\,-16\,G_0(x)\,G_{-x,1}(y)
+16\,G_{-1,0}(x)\,G_{-\frac{1}{x}}(y)-16\,G_{-1,0}(x)\,G_{-{x}}(y)-24\,\zeta_3\\
&\,-8\,G_0(y)\,G_{0,0}(x)-8\,G_{0,0}(x)\,G_{-\frac{1}{x}}(y)+8\,G_{0,0}(x)\,G_{-x}(y)+8\,G_{-\frac{1}{x},0,0}(y)\\
&\,+16\,G_{-\frac{1}{x},0,1}(y)+8\,G_{-x,0,0}(y)+16\,G_{-x,0,1}(y)-16\,G_{-1,0,0}(y)-32\,G_{-1,0,1}(y)\\
&\,-16\,G_{0,-1,0}(x)+8\,G_{0,0,0}(x)+\frac{20}{3}\pi^2\,G_{-\frac{1}{x}}(y)+4\pi^2\,G_{-x}(y)-\frac{32}{3}\pi^2\,G_{-1}(y)\,.
\esp\eeq
The function $Q(x,y)$ is related to exactly the same square root defined in eqs.~\eqref{eq:dxfbar} and~\eqref{eq:Delta_def}:
\beq
Q(x,y) = \sqrt{\frac{(x+y)(1+xy)}{x^2 y+xy^2-4 x y+x+y}} = \frac{\sqrt{\Delta(x,y)}}{x^2 y+xy^2-4 x y+x+y}\,,
\eeq
and the initial condition is $\widetilde{B}(x=1,y=1) = 0$. Following exactly the same steps as in section~\ref{sec:bhabha_empls}, we find the following very compact expression:
\begin{align}
\widetilde{B}(x,y)&\, = 16 \log \frac{-t}{m^2}\, \Big[\cEf{-1&1&1}{0&1+{1}/{y}&1}{\bar{x}}{\vec{a}}-\cEf{-1&1&1}{0&1+y&1}{\bar{x}}{\vec{a}}\\
\nonumber&\,+\cEf{-1&1&1}{\infty &1&1}{\bar{x}}{\vec{a}}+ \cEf{-1&1&1}{1&1&1}{\bar{x}}{\vec{a}}+\zeta_2\,\cEf{-1}{\infty }{\bar{x}}{\vec{a}}+\zeta_2\,\cEf{-1}{1}{\bar{x}}{\vec{a}}\Big]\\
\nonumber&\,-8\left(8 \zeta _2+4 \text{Li}_2(y)+ \log ^2y\right)\! \left[\cEf{-1&1}{0&1+{1}/{y}}{\bar{x}}{\vec{a}}+\cEf{-1&1}{0&1+y}{\bar{x}}{\vec{a}}-\cEf{-1&1}{0&1}{\bar{x}}{\vec{a}}\right]\\
\nonumber&\,-32\, \zeta _2\, \left[\cEf{-1&1}{\infty &1}{\bar{x}}{\vec{a}}- \cEf{-1&1}{1&1}{\bar{x}}{\vec{a}}\right]+16\, \cEf{-1&1&1&1}{0&1+{1}/{y}&1&1}{\bar{x}}{\vec{a}}\\
\nonumber&\,-32\, \cEf{-1&1&1&1}{0&1+{1}/{y}&2&1}{\bar{x}}{\vec{a}}-16\, \cEf{-1&1&1&1}{0&1+y&1&1}{\bar{x}}{\vec{a}}+32\, \cEf{-1&1&1&1}{0&1+y&2&1}{\bar{x}}{\vec{a}}\\
\nonumber&\,+16\, \cEf{-1&1&1&1}{\infty &0&1&1}{\bar{x}}{\vec{a}}-24\, \cEf{-1&1&1&1}{\infty &1&1&1}{\bar{x}}{\vec{a}}-32\, \cEf{-1&1&1&1}{\infty &1&2&1}{\bar{x}}{\vec{a}}\\
\nonumber&\,+16\, \cEf{-1&1&1&1}{1&0&1&1}{\bar{x}}{\vec{a}}+40\, \cEf{-1&1&1&1}{1&1&1&1}{\bar{x}}{\vec{a}}-32\, \cEf{-1&1&1&1}{1&1&2&1}{\bar{x}}{\vec{a}}\\
\nonumber&\,+\frac{4}{3}\big(12 \text{Li}_3(y)+24 \zeta _2 \log y+ \log ^3y\big) \left[\cEf{-1}{\infty }{\bar{x}}{\vec{a}}+\cEf{-1}{1}{\bar{x}}{\vec{a}}\right]\\
\nonumber&\,+64 \zeta _4-32 \zeta _2 \text{Li}_2(y)+16 \text{Li}_4(y)+8 \zeta _2 \log ^2y+\frac{1}{3}\log ^4y
\,.
\end{align}
We have again checked our result numerically. Note that again we find a pure function of uniform weight four.



\section{Conclusions}
\label{sec:conclusions}

In this paper we  considered the computation of the second family of planar master integrals relevant
for Bhabha scattering at two loops in QED including the full dependence on the electron mass. Our primary tool for their analytic study was the method of differential equations,
augmented by the choice of a canonical basis. We described how we obtained our canonical basis and showed that
four different square roots appear in the calculation of the 43 master integrals that comprise it. We also provided boundary conditions at $s=t=0$ for all master integrals. Together with the matrices defining the differential equations, this input is sufficient to produce high-precision numerical results for all master integrals and for all kinematic regions using the public code {\tt DiffExp}. We provide a Mathematica notebook that allows one to evaluate all master integrals numerically with {\tt DiffExp} as ancillary material with the arXiv submission.

We also considered the analytic solution of the differential equations. Interestingly, the four square roots never appear all at the same time, and the differential equations for 
all master integrals but one can be 
 solved in terms of MPLs by rationalising three of the four square roots 
by suitable changes of variables. For the contributions up to weight three, this procedure leads to analytic results which we present with the arXiv submission.
While conceptually straightforward, we find that this procedure generates rather involved analytical expressions for the weight four part of the master integrals. The analytic results are available in electronic form from the authors upon request.

In the last part of the paper, we focused on the analytic computation of the remaining master integral, 
whose canonical differential equations
contain three different square roots which cannot be rationalised at the same time.
An analytic calculation
in terms of MPLs cannot be easily obtained by direct integration of the differential equations due to the non-rationalisable square root. 
Instead, we show that compact analytic expressions can be obtained algorithmically 
in terms of elliptic multiple polylogarithms.
We applied this idea in detail to our problem and obtained in this way a very compact analytic expression
for this integral. We also showed that a similar compact expression can be obtained for another
planar integral relevant for two-loop Bhabha scattering, whose calculation in terms of (a lengthy combination of) MPLs 
had been considered some years ago.

This paper concludes the analytic calculation of the planar master integrals for Bhabha scattering at two loops. For the future, it would be interesting to complete also the computation of the non-planar two-loop family. Once this last step is achieved, we will have all the ingredient to obtain for the first time complete two-loop results in QED for one of the standard candle processes at an electron-positron collider.

\section*{Acknowledgments}

We are grateful to Martijn Hidding for help in applying {\tt DiffExp} and to Erik Panzer for suggesting the change of variable that renders the integral $f_{14}$ linearly reducible. We thank Marco Besier, Jake Bourjaily, Johannes Br\"odel, Falko Dulat and Brenda Penante for discussions. 
The work of V.S. was supported by the Russian Science Foundation, agreement no. 21-71-30003
(checking results with updated version of {\tt FIESTA})
and by 
the Ministry of Education and Science of the Russian Federation as part of the program of the Moscow Center 
for Fundamental and Applied Mathematics under agreement no. 075-15-2019-1621
(solving differential equations).

\appendix

\section{Canonical basis}
\label{app:canonical_basis}

In this appendix we show our choice of canonical basis. 
Note that we prefer to choose the square roots in such a form that that they are manifestly real in the Euclidean region $s,t<0$. 
This holds also for the boundary conditions which we presented in eq~\eqref{eq:bc_st0}.

\allowdisplaybreaks
\begin{align}
f_1&\,= \epsilon^2 F_{2,0,0,0,0,2,0,0,0}\,,\label{eq:canbas} \\
f_2&\,=-\epsilon ^2\frac{1}{2} \sqrt{-s}  \sqrt{4 m^2-s} F_{0,2,1,0,0,2,0,0,0}
-\epsilon ^2\sqrt{-s}  \sqrt{4 m^2-s} F_{0,2,2,0,0,1,0,0,0} \,,\nonumber \\
f_3&\,= - \epsilon ^2 s F_{0,2,1,0,0,2,0,0,0}\,,\nonumber \\
f_4&\,= -\frac{1}{2} \epsilon ^2 \sqrt{-t}  \sqrt{4 m^2-t} F_{0,0,0,0,1,2,2,0,0}
-\epsilon ^2 \sqrt{-t}  \sqrt{4 m^2-t} F_{0,0,0,0,2,1,2,0,0} \,,\nonumber \\
f_5&\,= - \epsilon ^2 t F_{0,0,0,0,1,2,2,0,0} \,,\nonumber \\
f_6&\,= -\epsilon ^2 m^2 F_{0,0,1,0,2,2,0,0,0}\, \nonumber \\
f_7&\,= -\epsilon ^3 \sqrt{-s}  \sqrt{4 m^2-s}  F_{0,1,1,0,1,2,0,0,0}\,, \nonumber \\
f_8&\,=  -\epsilon ^3 \sqrt{-t}   \sqrt{4 m^2-t}  F_{0,0,1,0,1,2,1,0,0}\,,\nonumber \\
f_9&\,=  -\epsilon ^2 m^2  F_{2,0,0,0,0,2,1,0,0}\,,\nonumber \\
f_{10}&\,=  -\epsilon ^2 m^2  F_{2,0,0,0,0,2,1,0,0}\,,\nonumber \\
f_{11}&\,= -\epsilon ^3 \sqrt{-s}  \sqrt{4 m^2-s}  F_{0,1,1,0,0,2,1,0,0}\,,\nonumber \\ 
f_{12}&\,= 2 \epsilon ^3 t  F_{0,1,0,0,1,1,2,0,0}- \epsilon ^2 t F_{0,1,0,-1,1,2,2,0,0}\,,\nonumber \\ 
f_{13}&\,=-\epsilon ^3 \sqrt{-t}   \sqrt{4 m^2-t} F_{0,1,0,0,1,1,2,0,0} \,,\nonumber \\ 
f_{14}&\,= -\epsilon ^4  \sqrt{-s-t}  \sqrt{4 m^2-s-t} F_{0,1,1,0,1,1,1,0,0}  \,,\nonumber \\ 
f_{15}&\,= \frac{1}{2} \epsilon ^3 \sqrt{-t} \sqrt{4 m^2-t} \left(2 m^2 F_{0,1,1,0,1,2,1,0,0}
+2 m^2 F_{0,2,1,0,1,1,1,0,0}-s F_{0,1,1,0,1,2,1,0,0}\right)   \,,\nonumber \\ 
f_{16}&\,= \epsilon ^3 \sqrt{-s} \sqrt{-t} \sqrt{4 m^2-s} \sqrt{4 m^2-t} F_{0,1,1,0,1,2,1,0,0} \,,\nonumber \\ 
f_{17}&\,=  \frac{1}{2} \epsilon ^3 \sqrt{-s}  \sqrt{4 m^2-s} \left(2 m^2
   F_{0,1,1,0,1,1,2,0,0}+2 m^2 F_{0,1,1,0,1,2,1,0,0}-t F_{0,1,1,0,1,2,1,0,0}\right)\,,\nonumber \\ 
f_{18}&\,= \frac{\epsilon ^3  m^2 \left(4 m^2-t\right) }{2 \left(4
   m^2-s-t\right)} F_{0,0,1,0,1,2,1,0,0}  
 -\frac{\epsilon ^3 \left(4 m^2-t\right) \left(2 m^2-s-t\right)}{2 \left(4 m^2-s-t\right)} F_{0,1,0,0,1,1,2,0,0} \nonumber \\
&\,+\frac{\epsilon ^3 m^2 s \left(4 m^2-s\right) }{t \left(4m^2-s-t\right)} F_{0,1,1,0,0,2,1,0,0}  
 -\epsilon ^3  \left(\epsilon 10 m^2 
   -m^2-3 \epsilon s  -3  \epsilon t\right) F_{0,1,1,0,1,1,1,0,0} \nonumber \\
&\, -\frac{\epsilon ^3 m^2 (s+t) }{t} F_{0,2,1,0,1,1,1,0,-1}  
 +\frac{ \epsilon ^2 m^2\left(4 m^2-t\right)}{4 \left(4 m^2-s-t\right)} F_{0,0,0,0,1,2,2,0,0} \nonumber \\
&\, +\frac{ \epsilon ^3 m^2\left(8 m^4-4 m^2 s-6 m^2 t+s^2+2 s t+t^2\right) }{4 m^2-s-t} F_{0,2,1,0,1,1,1,0,0} \nonumber \\
&\,+\frac{\epsilon ^2 m^2 \left(4 m^2-t\right) }{2 \left(4 m^2-s-t\right)} F_{0,0,0,0,2,1,2,0,0}  
+\frac{\epsilon ^3 m^2 \left(8 m^4-4 m^2 s-2 m^2 t+s^2+s t\right)}{4 m^2-s-t} F_{0,1,1,0,1,1,2,0,0}\nonumber \\
&\,+\frac{\epsilon ^3 \left(8 m^4-4 m^2 s-2 m^2 t+s^2+s t\right) }{4 \left(4 m^2-s-t\right)}F_{0,1,1,0,1,2,0,0,0}\nonumber \\
&\,+\frac{\epsilon ^2 \left(8 m^4-4 m^2 s-2 m^2 t+s^2+s t\right)}{8 \left(4
   m^2-s-t\right)} F_{0,2,1,0,0,2,0,0,0} \nonumber \\
&\,+\frac{\epsilon ^2 \left(8 m^4-4 m^2 s-2 m^2 t+s^2+s t\right) }{4 \left(4 m^2-s-t\right)} F_{0,2,2,0,0,1,0,0,0} \nonumber\\
&\,+\frac{1}{2} \epsilon ^3 \left(8 m^4-2 m^2 s-2 m^2 t+s t\right) F_{0,1,1,0,1,2,1,0,0}\,, \nonumber\\ 
f_{19}&\,= -\epsilon ^2 s  F_{2,0,2,1,0,0,0,0,0} \,, \nonumber \\
%
f_{20}&\,= \epsilon ^2 \sqrt{-s} s \sqrt{4 m^2-s} F_{2,1,2,1,0,0,0,0,0} \,,\nonumber \\ 
f_{21}&\,=  -\epsilon ^2 \sqrt{-s}  \sqrt{4 m^2-s}  F_{2,1,0,0,0,2,0,0,0} \,,\nonumber \\ 
f_{22}&\,=  -\frac{1}{2} \epsilon ^3 s F_{0,0,1,1,2,1,0,0,0}-\epsilon ^2 s F_{0,0,2,1,1,2,0,-1,0} \,,\nonumber \\ 
f_{23}&\,=  -\epsilon ^3 \sqrt{-s}   \sqrt{4 m^2-s}  F_{0,0,1,1,2,1,0,0,0} \,,\nonumber \\ 
f_{24}&\,=  -\epsilon ^3 \sqrt{-s}  \sqrt{4 m^2-s}  F_{0,0,1,1,1,2,0,0,0} \,,\nonumber \\ 
f_{25}&\,=  -\epsilon ^3 \sqrt{-s}  \sqrt{4 m^2-s}  F_{2,0,1,1,0,0,1,0,0} \,,\nonumber \\
f_{26}&\,=  \epsilon ^3 s  F_{1,1,0,0,0,2,1,0,0}-\epsilon ^2 s F_{2,1,0,0,0,2,1,0,-1} \,,\nonumber \\
f_{27}&\,=  - \epsilon ^3 \sqrt{-s} \sqrt{4 m^2-s} F_{1,1,0,0,0,2,1,0,0} \,,\nonumber \\
f_{28}&\,= \epsilon ^3 s F_{1,1,1,1,0,1,0,0,0}-2 s \epsilon ^4 F_{1,1,1,1,0,1,0,0,0}  \,,\nonumber \\
f_{29}&\,= - \epsilon ^4 \sqrt{-s} \sqrt{4 m^2-s} F_{1,0,1,1,1,1,0,0,0}  \,,\nonumber \\
f_{30}&\,=  \epsilon ^4 s^2 F_{1,1,1,1,1,1,0,0,0}-4\epsilon ^4  m^2 s F_{1,1,1,1,1,1,0,0,0} \,,\nonumber \\
f_{31}&\,=  \epsilon ^3 s^2 F_{2,1,1,1,0,0,1,0,0}-4 \epsilon ^3 m^2 s F_{2,1,1,1,0,0,1,0,0} \,,\nonumber \\
f_{32}&\,=  \epsilon ^4 \sqrt{-s} s  \sqrt{4 m^2-s} F_{1,1,1,1,0,1,1,0,0}\, \nonumber \\
f_{33}&\,=  -\epsilon ^3 \sqrt{-s}  \sqrt{4 m^2-s} F_{1,1,0,0,1,2,1,0,-1}
-\epsilon ^3 \sqrt{-s} t \sqrt{4 m^2-s} F_{1,1,0,0,1,2,1,0,0} \,,\nonumber \\
f_{34}&\,=  \epsilon ^3 \sqrt{-s} \sqrt{-t} \sqrt{4 m^2-s} \sqrt{4 m^2-t} F_{1,1,0,0,1,2,1,0,0} \,,\nonumber \\
f_{35}&\,=  -\epsilon ^3 \sqrt{-s}  \sqrt{4 m^2-s} F_{0,0,1,1,1,2,1,-1,0} \,,\nonumber \\
f_{36}&\,= \epsilon ^3 s \sqrt{-t}  \sqrt{4 m^2-t} F_{0,0,1,1,1,2,1,0,0}
+\epsilon ^3 s \sqrt{-t} \sqrt{4 m^2-t} F_{0,0,1,1,2,1,1,0,0}  \,,\nonumber \\
f_{37}&\,=  \epsilon ^3 \sqrt{-s} \sqrt{4 m^6-s \left(m^2-t\right)^2} F_{0,0,1,1,1,2,1,0,0} \,,\nonumber \\
f_{38}&\,=\epsilon ^4 s \sqrt{-t} \sqrt{4 m^2-t} F_{1,0,1,1,1,1,1,0,0} \,,\nonumber \\
f_{39}&\,= - \epsilon ^4 \sqrt{-s}\sqrt{4 m^2-s}\left(m^2
   F_{1,0,1,1,1,1,1,0,0}-F_{0,1,1,0,1,1,1,0,0}+F_{1,0,1,1,1,1,1,-1,0} \right)\,,\nonumber \\
f_{40}&\,= -\epsilon ^4 s \sqrt{-s} \sqrt{-t} \sqrt{4 m^2-s} \sqrt{4 m^2-t} F_{1,1,1,1,1,1,1,0,0}\,,\nonumber \\
f_{41}&\,= -\epsilon ^4  s \sqrt{-s}   \sqrt{4 m^2-s} (F_{1,1,1,1,1,1,1,0,-1}+t F_{1,1,1,1,1,1,1,0,0})\,,\nonumber \\
f_{42}&\,= \epsilon ^4\left( 4 m^4 s  F_{1,1,1,1,1,1,1,0,0}-m^2 s^2   F_{1,1,1,1,1,1,1,0,0} \right. \nonumber \\
&\,\left. +4 m^2 s  F_{1,1,1,1,1,1,1,-1,0}-s^2   F_{1,1,1,1,1,1,1,-1,0}\right) \,,\nonumber \\
%
f_{43}&\,= 
\epsilon ^4 m^2 s  F_{1,0,1,1,1,1,1,0,0}-\epsilon ^4 m^2 s F_{1,1,1,1,1,1,1,0,-1}
-\epsilon ^3 m^2 s F_{0,1,1,0,1,1,2,0,0} \nonumber \\
&\,-\epsilon ^3 m^2 s F_{0,1,1,0,1,2,1,0,0}
+\epsilon ^4 \frac{1}{2} s^2 t F_{1,1,1,1,1,1,1,0,0}
-\epsilon ^4 \frac{1}{2} s^2 F_{1,1,1,1,0,1,1,0,0} \nonumber\\
&\,+\epsilon ^4 \frac{1}{2} s^2 F_{1,1,1,1,1,1,1,0,-1}
+\frac{1}{2}\epsilon ^3  s t F_{0,1,1,0,1,2,1,0,0}-\epsilon ^3 s t F_{1,1,0,0,1,2,1,0,0}\,, \nonumber \\
&\,+\epsilon ^3 s t F_{2,1,1,1,0,0,1,0,0}
+\epsilon ^4 s F_{0,1,1,0,1,1,1,0,0} 
 ,-\epsilon ^4 s F_{1,0,1,1,1,1,0,0,0}+\epsilon ^4 s F_{1,0,1,1,1,1,1,-1,0}  \nonumber\\
&\,-\epsilon ^4 s F_{1,1,1,1,1,1,1,-1,-1}-\frac{1}{2} \epsilon ^3 s F_{0,1,1,0,0,2,1,0,0} 
+\frac{1}{4} \epsilon ^3 s F_{0,1,1,0,1,2,0,0,0}+\epsilon ^3 s F_{1,1,0,0,0,2,1,0,0} \nonumber\\
&\,-\epsilon ^3 s F_{1,1,0,0,1,2,1,0,-1}+\frac{1}{2} \epsilon ^3 s F_{1,1,1,1,0,1,0,0,0}
-\epsilon ^2 \frac{1}{8} s F_{0,2,1,0,0,2,0,0,0}
-\epsilon ^2 \frac{1}{4} s F_{0,2,2,0,0,1,0,0,0}
\,.\nonumber
\end{align}

 
 \section{Elliptic multiple polylogarithms}
 \label{app:eMPLs}
 
 In this appendix we collect some technical material related to eMPLs not reviewed in section~\ref{sec:eMPL_algo}.
 
 In the case where all the indices $A_i = \left(\begin{smallmatrix} n_i\\ c_i\end{smallmatrix}\right)$ are equal to $\left(\begin{smallmatrix} \pm 1\\ 0\end{smallmatrix}\right)$, the integral in eq.~\eqref{eq:cE4_def} is divergent and requires a special treatment
  and we define instead,
\begin{align}\label{eq:eMPL_reg}
\cE_4(A_1\ldots A_k;x;\vec a) = \frac{1}{k!}\log^kx + &\sum_{l=0}^k\sum_{m=1}^l\sum_{\sigma}\frac{(-1)^{l+m}}{(k-l)!}\log^{k-l}x\\
\nonumber&\times\cE_4^{\textrm{R}}\left(A^{(m)}_{\sigma(1)}\ldots A^{(m)}_{\sigma(m-1)} A^{(m)}_{\sigma(m+1)}\ldots A^{(m)}_{\sigma(l)}\Big|A_m;x;\vec a\right)\,,
\end{align}
with $A_i^{(m)} = A_i$ if $i<m$ and $A_i^{(m)} = \left(\begin{smallmatrix}1\\0\end{smallmatrix}\right)$ otherwise. The third sum runs over all shuffles $\sigma$ of $\left(A^{(m)}_{1}\ldots A^{(m)}_{m-1}\right)$ and $\left(A^{(m)}_{m+1}\ldots A^{(m)}_{l}\right)$, i.e., over all permutations of their union that preserve the relative ordering within each of the two lists. The $\cE_4^{\textrm{R}}$ are iterated integrals with suitable subtractions to render the integrations finite,
\beq
\!\!\!\cE_4^{\textrm{R}}\left(\begin{smallmatrix} n_1&\ldots & n_k \\ 0 &\ldots &0\end{smallmatrix}|\begin{smallmatrix} n_a \\ 0 \end{smallmatrix};x;\vec a\right) 
= \int_0^xdt_1\Psi_{n_1}(0,t_1)\int_{0}^{t_1}\ldots \int_{0}^{t_{k-1}}dt_k\left(\Psi_{n_a}(0,t_k) - \Psi_{1}(0,t_k)\right)\,.
\eeq
The rather ad-hoc looking form of eq.~\eqref{eq:eMPL_reg} is determined essentially uniquely by requiring the regularised eMPLs to share the same algebraic and differential properties as their convergent analogues. We refer to ref.~\cite{Broedel:2018qkq} for a detailed discussion.

The functions $Z_4(x,\vec a)$ and $G_{\ast}(\vec a)$ that appear in eq.~\eqref{eq:pure_psi1} can be expressed in terms of incomplete elliptic integrals of the first and second kind~\cite{Broedel:2017kkb,Broedel:2018qkq}:
\beq\bsp
\textrm{F}(x|\lambda) &\,= \int_0^x\frac{dt}{\sqrt{(1-t^2)(1-\lambda t^2)}}\,,\\
\textrm{E}(x|\lambda) &\,= \int_0^xdt\,\sqrt{\frac{1-\lambda t^2}{1-t^2}}\,.
\esp\eeq

\bibliography{bha2}

\end{document}